\documentclass[fleqn,usenatbib]{mnras}

\usepackage[T1]{fontenc}
\usepackage{ae,aecompl}
\usepackage{soul}

\usepackage{graphicx}	% Including figure files
\usepackage{amsmath}	% Advanced maths commands
\usepackage{amssymb}	% Extra maths symbols
\usepackage{threeparttable}
\newcommand{\lSect}[1]{{\label{sec:#1}}}
\newcommand{\lFig}[1]{{\label{fig:#1}}}
\newcommand{\lEq}[1]{{\label{eq:#1}}}
\newcommand{\lTab}[1]{{\label{tab:#1}}}

\def\gtaprx {\lower .1ex\hbox{\rlap{\raise .6ex\hbox{\hskip .3ex
        {\ifmmode{\scriptscriptstyle >}\else
                {$\scriptscriptstyle >$}\fi}}}
        \kern -.4ex{\ifmmode{\scriptscriptstyle \sim}\else
                {$\scriptscriptstyle\sim$}\fi}}}
\def\ltaprx {\lower .1ex\hbox{\rlap{\raise .6ex\hbox{\hskip .3ex
        {\ifmmode{\scriptscriptstyle <}\else
                {$\scriptscriptstyle <$}\fi}}}
        \kern -.4ex{\ifmmode{\scriptscriptstyle \sim}\else
                {$\scriptscriptstyle\sim$}\fi}}}
\newcommand{\FIGFF}[2]{{\ref{fig:#2}{#1}}}

\newcommand{\FIG}[2]{{Fig.~\FIGFF{#1}{#2}}}
\newcommand{\Fig}[1]{{\FIG{}{#1}}}

\newcommand{\Sectff}[1]{{\ref{sec:#1}}}
\newcommand{\Sect}[1]{{\S\Sectff{#1}}}

\newcommand{\Eqref}[1]{{\ref{eq:#1}}}

\newcommand{\eqff}[1]{{\Eqref{#1}}}

\newcommand{\eq}[1]{{equation~\eqff{#1}}}

\newcommand{\Leg}[1]{{\textit{#1}}}
\newcommand{\Msun}{\ensuremath{\mathrm{M}_\odot}}
\newcommand{\Lsun}{\ensuremath{\mathrm{L}_\odot}}

\newcommand{\Tab}[1]{{Table \ref{tab:#1}}}

\newcommand{\KEPLER}{\ensuremath{\mathrm{\texttt{KEPLER}}}}
\newcommand{\FRANEC}{\ensuremath{\mathrm{\texttt{FRANEC}}}}

\newcommand{\MESA}{\ensuremath{\mathrm{\texttt{MESA}}}}

\newcommand{\GRD}{\ensuremath{\mathrm{\texttt{GR1D}}}}
\newcommand{\PHOTB}{\ensuremath{\mathrm{\texttt{PHOTB}}}}

\hypersetup{draft}

\title[Missing Red Supergiants and Carbon Burning]{Missing Red Supergiants and Carbon Burning}

\author[T. Sukhbold \& S. Adams]{
Tuguldur Sukhbold,$^{1,2}$\thanks{NASA Hubble Fellow, e-mail: tuguldur.s@gmail.com},
\& Scott Adams$^{3}$
\\
$^{1}$Department of Astronomy, The Ohio State University, 140 West 18th Avenue, Columbus OH 43210\\
$^{2}$Center for Cosmology and AstroParticle Physics, The Ohio State University, 191 W. Woodruff Avenue, Columbus OH 43210\\
$^{3}$Cahill Center for Astrophysics, California Institute of Technology, Pasadena, CA 91125
}

\date{Accepted XXX. Received YYY; in original form ZZZ}

\pubyear{2019}

\begin{document}
\label{firstpage}
\pagerange{\pageref{firstpage}--\pageref{lastpage}}
\maketitle
\begin{abstract}
Recent studies on direct imaging of Type II core-collapse supernova progenitors indicate a possible threshold around $M_{\rm ZAMS}\sim 16-20$ \Msun, where red supergiants with larger birth masses do not appear to result in supernova explosions and instead implode directly into a black hole. In this study we argue that it is not a coincidence that this threshold closely matches the critical transition of central Carbon burning in massive stars from the convective to radiative regime. In lighter stars, Carbon burns convectively in the center and result in compact final presupernova cores that are likely to result in explosions, while in heavier stars after the transition, it burns as a radiative flame and the stellar cores become significantly harder to explode. Using the \KEPLER\ code we demonstrate the sensitivity of this transition to the rate of $^{12}$C$(\alpha,\gamma)^{16}$O reaction and the overshoot mixing efficiency, and we argue that the upper mass limit of exploding red supergiants could be employed to constrain uncertain input physics of massive stellar evolution calculations. The initial mass corresponding to the central Carbon burning transition range from 14 to 26 \Msun\ in recently published models from various groups and codes, and only a few are in agreement with the estimates inferred from direct imaging studies.
\end{abstract}


%%%%%%%%%%%%%%%%%%%%%%%%%%%%%%%%%%%%%%%%%%%%%%%%%%%%%%%%%%%%%%%%%%%%%%%%%%%
\section{Introduction}
\lSect{intro}

It would be fair to say that most of the massive stars end their lives as red supergiants (RSG), unless they experience complicated binary interaction or extreme mass loss, rotation or they originate from a low metallicity environment. At the end of their lives, the iron-cores will inevitably collapse, which could either lead to a bright supernova explosion or to a rather quiet implosion into a stellar mass black hole. Given the connection between RSG stars and Type II core-collapse superpernovae, it is only natural to ask - do all RSG die in a supernova? if not, which ones end up exploding and which ones end up imploding? Both observational and theoretical studies from the past two decades point to intriguing and complicated answers to these questions.

\begin{figure*}%{r}{0.5\textwidth}
\includegraphics[width=.49\textwidth]{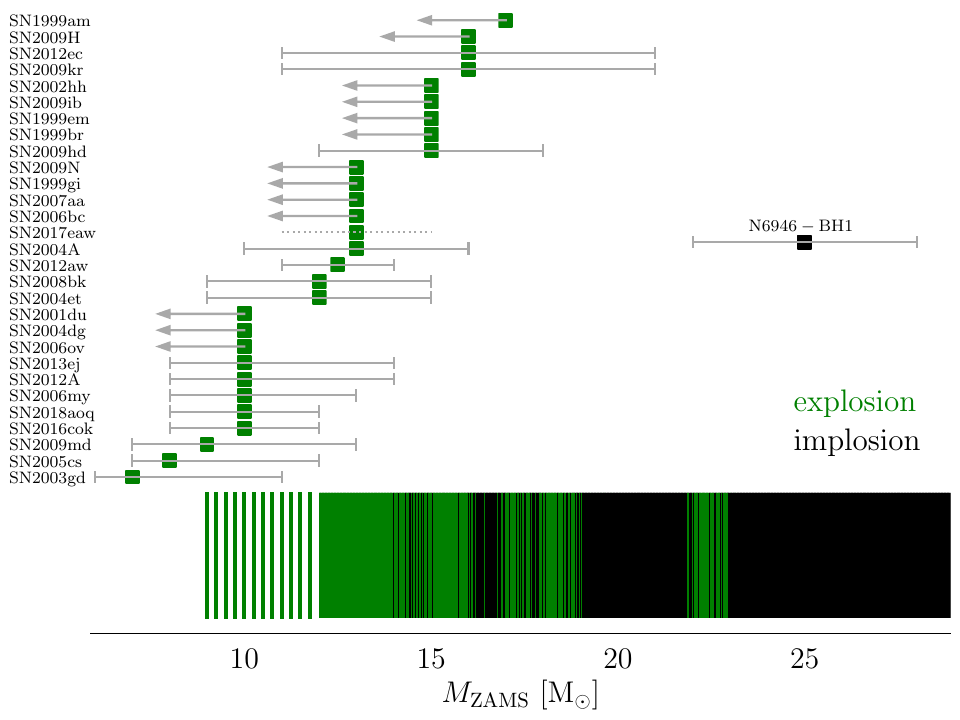}
\includegraphics[width=.49\textwidth]{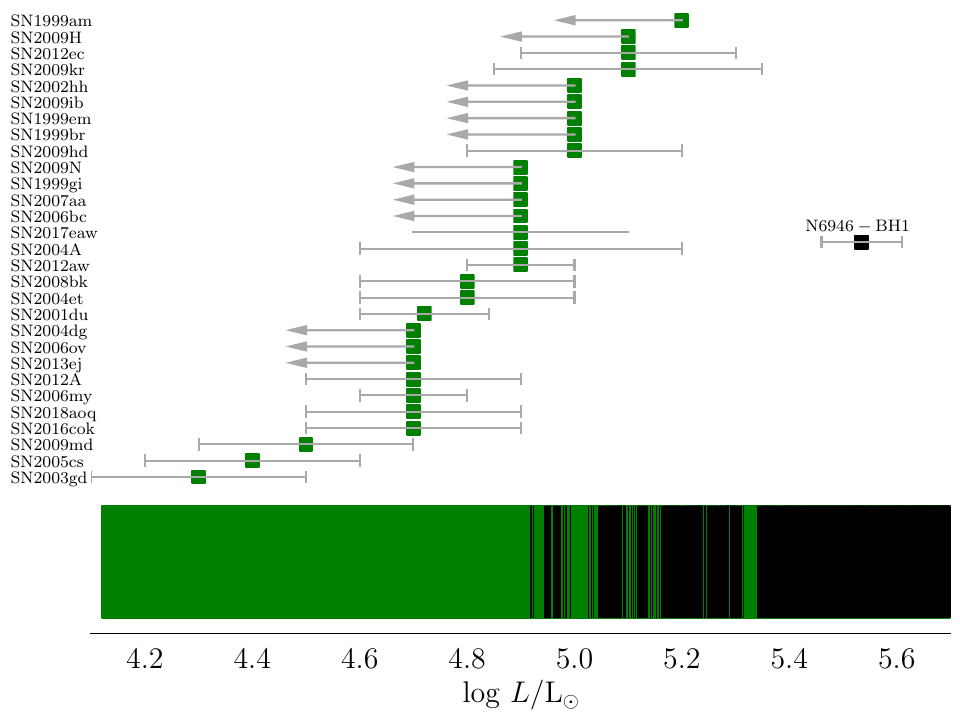}
\caption{(Horizontal bars, arrows:) Compilation of currently known progenitor initial mass and luminosity estimates from direct imaging studies. Arrows indicate upper limits and dotted line indicates a soft limit. Stars that resulted in Type II supernova are shown with green squares, which consists from the data compiled by \citet{Sma15} along with the estimates for SN2016cok \citep{Koc17}, SN2017eaw \citep{Kil18}, and SN2018aoq \citep{One18}. The values for SN2012aw were updated by the estimates from \citet{Fra16}. The only detection of direction implosion candidate \citep{Ada17} is shown in black with the updated distance of $7.72\pm0.32$ Mpc to N6946BH-1 \citep{Eld19}. The vertical bars show the final outcomes for non-rotating solar metallicity progenitors \citep{Suk18,Suk16} based on a sample calibrated neutrino-driven explosion model \citep{Ert16}. The comparison is made on the initial mass space (left), and on the luminosity space (right). Although the correspondence between data and models does not change significantly, the observations more directly infer the luminosity of the progenitor star, which is largely set by the luminosity of the embedded Helium-core. The Helium core mass, $M_{\alpha}$, and its luminosity at presupernova time, $L$, roughly scale as $\log (L/\Lsun) \sim 5.18\times(M_\alpha/6\Msun)^{0.12}$ \citep{Suk18}.\lFig{map}}
\end{figure*}

From the observational side, direct imaging studies are providing unique opportunities to decode the relationship between the birth properties and final fates of massive stars. On one end, \citet{Koc08} pointed out the absence of high mass progenitors, and proposed that a systematic search of disappearing stars could reveal whether most luminous RSG experience direct implosion into a black hole. \citet{Ada17} and \citet{Ger15} reported the first confirmed detection after surveying nearby galaxies for seven years, in which they found a red source consistent with a $M_{\rm ZAMS}\sim 25\ \Msun$ RSG star disappearing after a long-duration weak transient that resembled theoretical predictions for neutrino-mediated mass loss \citep{Lov13}.

On the other end, compiling initial mass estimates of 20 Type-IIP supernova progenitors into a volume limited sample, \citet{Sma09} found an upper-limit of $M_{\rm ZAMS}\sim 16-20$ \Msun. This is significantly smaller than the maximum initial mass for stars expected to end their lives as RSG \citep[thought to be $\sim 25-30\ \Msun$, e.g.,][]{Lev09,Mas00}, implying that most of the RSG above this threshold do not appear to die in a supernova. This result has continued to hold with additional progenitor initial mass estimates \citep[see, e.g., the review by][]{Sma15}, and is now widely known as the ``missing RSG problem''.

Potential biases due to selection effects, dust and luminosity estimates in direct imaging studies were extensively discussed in \citet{Sma15}, however, \citet{Dav18} argue that the progenitor luminosities are being underestimated, and that the maximum limit could be as high as $M_{\rm ZAMS}\sim25-35$ \Msun\ (though see discussion in \Sect{disc}). Many indirect  progenitor mass estimates are broadly consistent with the threshold mass from direct imaging results \citep[e.g.,][]{Jen14,Jer14,Val16}, however, some are inconsistent or inconclusive \citep[e.g.,][]{Kat18}. While direct imaging has not detected any RSG progenitor above $M_{\rm ZAMS}> 20\ \Msun$, there is some tentative evidence to suggest that not all RSG with larger initial mass disappear without a supernova. For example, through simple radiation-hydrodynamical modeling \citet{Das18} estimates an initial mass of 24--26 \Msun\ for the progenitor of SN2015ba, however, it should be noted that this type of light-curve modeling approach is not robust and does not provide a unique solution \citep[e.g.,][]{Des19}. Encapsulating all the evidence at hand (\Fig{map}, horizontal bars), recent observational studies indicate a threshold in both the initial mass and luminosity space below which most or all RSGs result in a supernova ($M_{\rm ZAMS}\lesssim 20\ \Msun$ or $\log L/\Lsun \lesssim 5.2$) and above which most RSGs disappear without a bright explosion ($M_{\rm ZAMS}\gtrsim 20\ \Msun$ or $\log L/\Lsun \gtrsim 5.2$).

From the theoretical side, we currently have two distinct scenarios. One set of solutions propose that the missing luminous RSG progenitors could be explained if RSG stars from higher initial masses can transform into blue stars before death. In some studies this is achieved by enhancing mass loss at higher initial masses. For instance, models published in \citet{Gro13} employ enhanced mass loss for stars above $M_{\rm ZAMS}\gtrsim 18$ \Msun\ on supra-Eddington luminosity considerations, so that the stars which would have died as luminous RSG instead died as compact blue stars (either luminous blue variables or Wolf-Rayet). In other studies this results emerges as a consequence of employing a strong mass loss rate, determined either empirically or theoretically. For example, \citet{Yoo10} explored the effect of ``superwind'' due to pulsations driven by the partial ionization of Hydrogen in the envelope \citep[e.g.,][]{Heg97}, and argued that the maximum mass for the star to retain Hydrogen envelope is close to 19--20 \Msun. \citet{Chi13} and \citet{Lim18} employed one of the most powerful mass loss prescriptions \citep{van05} and find that models with $M_{\rm ZAMS}\gtrsim 18$ \Msun\ die as Wolf-Rayet stars. However, such a scenario is in tension with the claimed detection of an imploding luminous RSG \citep{Ada17}. The required powerful mass loss rate may be in tension with other observational constraints \citep[e.g.,][]{Bea18}, and furthermore, as pointed out by \citet{Sma15}, it may also be inconsistent with the observations of Type-Ib/c supernovae.

The other scenario argues that massive stars that die as luminous RSG are intrinsically harder to explode, and their collapse generally leads to an implosion into a stellar mass black hole. While the idea was hypothesized since the 80s for various reasons (e.g., \citealp{Twa87}, \citealp{Mae92} and \citealp{Bro94} invoked to due to nucleosynthesis considerations), its modern realization emerged from decades of work on the advanced stage evolution of massive stars and the mechanism of core-collapse supernova explosions. It has long been known that the final fate of a massive star is strongly tied to the progenitor star's core structure just before its death \citep{Bur95} and that this structure varies non-monotonically with initial mass \citep{Wea93,Tim96}. A number of one-dimensional, calibrated, neutrino-driven explosion studies from the past several years \citep{Ott11,Ugl12,Hor14,Pej15,Ert16,Suk16,Mul16,Ebi19} highlighted the complicated explosion landscape, which is largely dictated by the non-monotonically varying final core structures of massive stars.

The emerging picture (\Fig{map}, vertical bars) indicates that most of the massive stars up to about $M_{\rm ZAMS}\sim20$ \Msun\ have compact cores that are easier to blow up, while more massive stars retain extended core structures that tend to implode\footnote{By compact final core structure we refer to smaller mass iron-core surrounded by a steeply declining density gradient, i.e. the structure described by a smaller value of the compactness parameter $\xi$ \citep{Ott11}, and smaller values of $M_4$ and $\mu_4$ \citep{Ert16}. By extended presupernova core structure we mean the opposite.}. However, there is no clean threshold that separates the two outcomes, instead there are narrow ranges of successful explosions above 20 \Msun, and implosions below as well. This scenario is not only consistent with direct progenitor imaging results, but it is also in a broad agreement with nucleosynthesis and light curves \citep{Suk16,Bro13}, compact object mass distributions \citep{Koc14,Koc15,Rai18}, and explosion energies and $^{56}$Ni masses \citep{Mul17}. 

The possible connection between the missing luminous RSG supernova progenitor stars and massive stellar cores becoming abruptly harder to blow up above $M_{\rm ZAMS}\sim20$ \Msun\ was first pointed out by \citet{Hor14}. However, the non-monotonically varying final core structures across the initial mass space is ultimately the consequence of massive stellar evolution. The systematic studies of presupernova evolution by \citet{Suk14} and \citet{Suk18} suggested that the interplay of convective burning episodes during the advanced stage evolution of the stellar core (from Carbon-burning until collapse) play a key role in determining the presupernova core structure, and thus the final outcome of its collapse. In particular, the relevant interplay for the missing RSG problem near $M_{\rm ZAMS}\sim20$ \Msun\ is between convective Carbon burning shells and the Oxygen burning convective core, which is largely driven by the critical transition of the central Carbon burning episode.

In this work, we argue that the observed initial mass threshold, separating the RSG stars that tend to die in a supernova explosion from those tend to disappear, is an indirect signature of central Carbon burning in massive stars transitioning from the convective to radiative regime. We also suggest that the observed threshold mass could be employed to constrain massive stellar evolution calculations. To this end, we first provide an overview on the physics of the critical Carbon burning transition, and we discuss its connection to explodability utilizing previously published calculations (\Sect{trans}). We then review the relevant uncertain input physics that determine the birth mass for this transition, provide sample sensitivity calculations, and survey the results seen in the literature (\Sect{constr}). We end by discussing the caveats of our arguments, and the potentially testable prediction based on the models (\Sect{disc}).

%%%%%%%%%%%%%%%%%%%%%%%%%%%%%%%%%%%%%%%%%%%%%%%%%%%%%%%%%%%%%%%%%%%%%%%%%%%
\section{Critical Transition of Carbon Burning and Explodability}
\lSect{trans}

Massive stars typically live for millions of years, yet their final fate is strongly dependent on the advanced stages of evolution - the evolution that takes place in its core during its final few thousand years. A key feature of this part of the evolution is the copious energy loss through neutrinos. The entire advanced stage evolution of the core can be characterized as a Kelvin-Helmholtz contraction driven by the neutrino losses, that gets temporarily interrupted by nuclear burning episodes of heavier fuels \citep{Woo02}.

Carbon ignites at the center when the temperature exceeds roughly $5\times10^8$ K, marking the onset of the advanced stage evolution. Stellar modelers have long noted that Carbon burns convectively in the cores of lower initial mass ($\lesssim20$ \Msun) presupernova stars, and at higher mass ($\gtrsim20$ \Msun) it burns radiatively, where the transition sets an important milestone in the evolution that drastically affects the final presupernova core structure and properties of the explosion \citep[e.g.,][]{Wea93,Tim96,Bro99}.

The key condition that determines the character of Carbon burning, whether in a convective episode or as a radiative flame, is the ratio of the energy generation rate from nuclear burning ($\epsilon_{\rm n}$) to the energy loss rate through neutrinos ($\epsilon_{\nu}$). Ignoring the flux carried by photons, the necessary condition for driving convection can be written as \citep[e.g.,][]{Bar94},
\begin{equation}
\frac{{\rm d}s}{{\rm d}t}=\frac{\epsilon_\nu}{T}\Big(\frac{\epsilon_{\rm n}}{\epsilon_\nu}-1\Big)>0,
\lEq{conv}
\end{equation}
where $s$, $T$ and $t$ are entropy density, temperature and time respectively. To drive convection we require a local increase in the entropy density, and since $\epsilon_\nu$ and $T$ are both positive the condition is really just $\epsilon_{\rm n}/\epsilon_{\nu}>1$. In lower mass presupernova stars, the rate of energy generation from central Carbon burning more than compensates for the neutrino losses, while at higher mass it never exceeds the loss rate.

\begin{figure*}%{r}{0.99\textwidth}
\includegraphics[width=.98\textwidth]{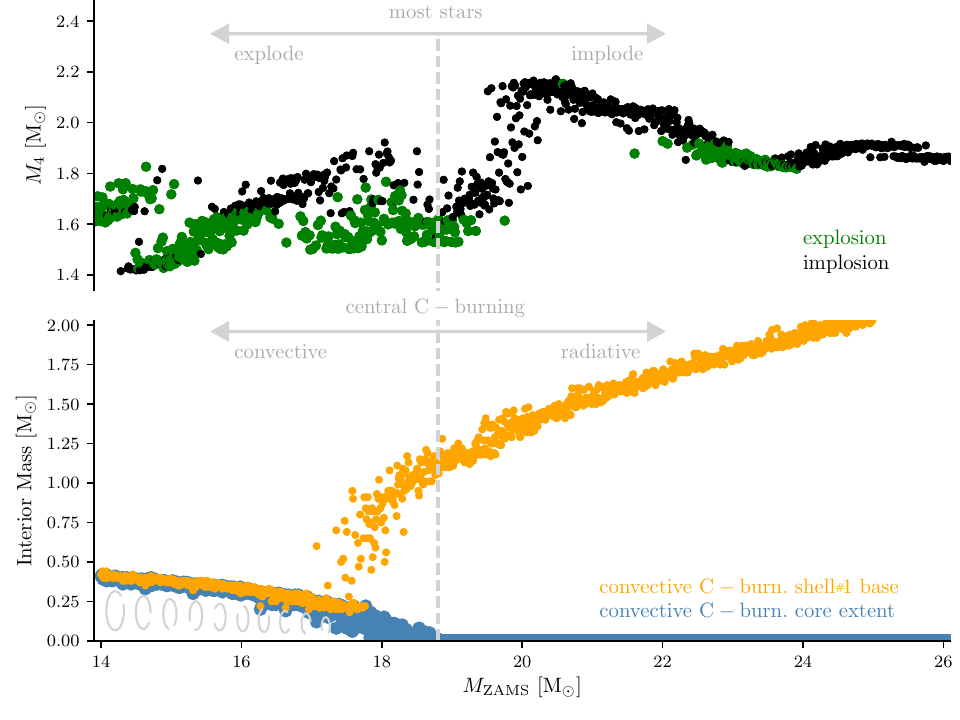}
\caption{(Top:) Lagrangian location in the \citet{Suk18} presupernova models where the entropy per baryon is 4$k_{\rm B}$ ($M_4$), which mostly tracks the location of the strongest Oxygen burning shell. Smaller values correspond to models with compact cores (see footnote 1) that are easier to explode, and larger values represent models with extended cores that are harder to explode. The final outcomes (explosion-green, implosion-black) are determined by the 2-parameter criterion of \citet{Ert16} with the N20 engine.
(Bottom:) Mass extent of the convective Carbon burning core (blue) and the base of first convective Carbon burning shell (orange) for each model. The central Carbon burning transitions from convective to radiative regime near $M_{\rm ZAMS}\approx 18.7 \Msun$. The innermost convective Carbon burning shell rapidly ``migrates'' outwards with increasing mass near this transition, which ultimately results in more extended core structures that are likely to implode. The transition mass separates the initial mass space where most stars are likely to explode (below) and implode (above). Assuming a Salpeter-IMF with $\alpha=-2.35$, 89\% of RSG between 9 \Msun\ and 18.7 \Msun\ explode, and 84\% between 18.7 \Msun\ and 27 \Msun\ implode. \lFig{f2}}
\end{figure*}

In the simplest sense, two key parameters determine the ratio $\epsilon_{\rm n}/\epsilon_{\nu}$ at the onset of central Carbon burning: the initial \textit{mass} and \textit{composition} of the Carbon-Oxygen (CO) core. Understanding the dependence of these parameters on the initial mass of the star helps to provide the reason behind the critical transition. The mass of the CO-core is straightforward. Hydrogen always burns in a massive convective core, whose extent effectively determines the mass of the resulting Helium core. Helium also burns in a massive convective core and its extent determines the embedded CO-core mass. Many calculations \citep[e.g.,][]{Woo02} have demonstrated that CO-core mass increases monotonically with initial mass until the entire envelope is removed through mass loss (i.e., until the maximum mass for single stars to die as RSG).

As for the composition, thanks to the convective central Helium burning, it is safe to assume that the evolution of the entire CO-core starts from a uniform mixture of mostly Carbon and Oxygen. The actual mass fraction of the available fuel at central Carbon ignition is determined by the competing reactions taking place during core Helium burning---$3\alpha$ vs. $^{12}$C$(\alpha,\gamma)^{16}$O \citep[e.g.,][]{Tur07}. Calculations have demonstrated that even at the lowest initial masses, the CO-core starts from a fairly Oxygen rich composition, which only gets further enriched with increasing initial mass (less available fuel). Since the density at which burning occurs is lower in higher mass stars (higher entropy), the rate of $3\alpha$ decreases as compared to the two-body reaction $^{12}$C$(\alpha,\gamma)^{16}$O. Therefore, the ratio $^{12}$C$/^{16}$O at the onset of Carbon ignition decreases with increasing initial mass. For the models in \citet{Suk18}, we find a central $^{12}$C mass fraction at the time of Carbon ignition of $X_{\rm C}(0)=$0.25, 0.21, 0.19, and  0.18 for initial masses of $M_{\rm ZAMS}=$ 12, 15, 20 and 25 \Msun, respectively.

With increasing initial mass, the star has a more massive CO-core and less available fuel for Carbon burning. How do these two aspects conspire to change the burning mode from convective to radiative? The answer to this question was provided more than two decades ago semi-analytically and numerically by \citet{Bar90} and \citet{Bar94}. Given the inaccessibility of these proceedings, we briefly reiterate their main argument here. The rates of energy generation from Carbon burning and loss through neutrinos can be written as
\begin{equation}
\epsilon_{\rm n} \sim X_{\rm C}^2 \rho T^{23} \quad {\rm and} \quad \epsilon_\nu \sim T^{12} \rho^{-1},
\lEq{eps}
\end{equation}
where $\rho$ and $T$ are density and temperature near the center. For an ideal, gas pressure dominated polytrope, the mass ($m$) scales as $m^2 \sim T^{3} \rho^{-1}$. Combining with \eq{eps}, the maximum of $\epsilon_{\rm n}/\epsilon_{\nu}$ scales as,
\begin{equation}
\frac{\epsilon_{\rm n}}{\epsilon_\nu}\Big|_{\rm max} \sim X_{\rm C}(0)^{1.4} \frac{{\rm d}T}{{\rm d}t}^{0.6}m^{-2.8}.
\end{equation}
Since the temperature does not change appreciably during fuel burning, this result shows that with an increasing mass and a decreasing amount of initial fuel, eventually there will be a mass above which the condition of \eq{conv} is no longer satisfied and the Carbon must burn radiatively. 

The transition sets the course for the advanced stage evolution so that the final presupernova structure becomes abruptly harder to blow up above this critical initial mass. Without convective central burning, the core effectively bypasses the long lasting neutrino-cooling phase, and leaves the core with much higher entropy. More specifically, as Carbon burns radiatively in the center, the flame propagates outward until it reaches a mass location where \eq{conv} is satisfied and the first convective Carbon burning shell is born. As pointed out in \citet{Suk14}, with increasing initial mass after the transition, the radiative burning travels further out in mass, which then pushes the base of the first Carbon burning shell further as well. This outward ``migration'' (with increasing initial mass) not only delays Oxygen ignition in the core but it also allows the convective core Oxygen burning episode to have larger mass extent. A massive Oxygen burning core results in a massive Silicon burning core, which ultimately leads to extended final presupernova structures that are hard to blow up.

The connection between explodability and central Carbon burning mode is illustrated in \Fig{f2}. Depicting the final core structures for 1200 solar metallicity non-rotating models \citep{Suk18} with initial masses between 14 and 26 \Msun, the top panel shows the mass shell in each presupernova star where entropy per baryon first exceeds $4 k_{\rm B}$ going outward ($M_4$). The final fates of each stars are determined through the 2-parameter criterion by \citet{Ert16} calibrated to one of their sample engines (N$20$). Since $M_4$ tracks the location of the strongest Oxygen burning shell \citep{Suk18}, it shows that models below about 20 \Msun\ have $M_4\lesssim 1.8$ \Msun\ \citep{Bro99} and result in compact cores that are typically easy to explode, while heavier models always have $M_4>1.8$ \Msun\ and have extended cores that are much more difficult to explode. Here note that the same qualitative result will persist if we instead use the Fe-core mass or compactness parameter \citep{Ott11}, or any other simple descriptor of the core structure, in combination with a weaker or stronger explosion engine model.

The bottom panel shows the mass extent of the convective Carbon burning core (if applicable) and the base of the innermost Carbon burning shell for each corresponding model. The extent of the convective Carbon burning core shrinks with increasing initial mass, as the CO-core mass grows and the available fuel at the onset of Carbon ignition decreases. For the adopted set of models, near $M_{\rm ZAMS}\approx 18.7$ \Msun\ the central Carbon burning mode transitions, and for more massive models it burns radiatively. As pointed out by \citet{Suk14}, the Carbon burning transitions to radiative regime at an off-center location at a slightly lower initial mass due to degeneracy, and therefore the base of the first Carbon burning shell shoots up slightly before the central burning experiences the transition. The rapid outward ``migration'' of the innermost Carbon burning shell delays the ignition of Oxygen in the core and makes it more massive. The increasing extent of the Oxygen burning core pushes the Oxygen burning shell outward, which eventually results in extended core structures with relatively high values of $M_4$.

\begin{figure}%{r}{0.5\textwidth}
\includegraphics[width=.48\textwidth]{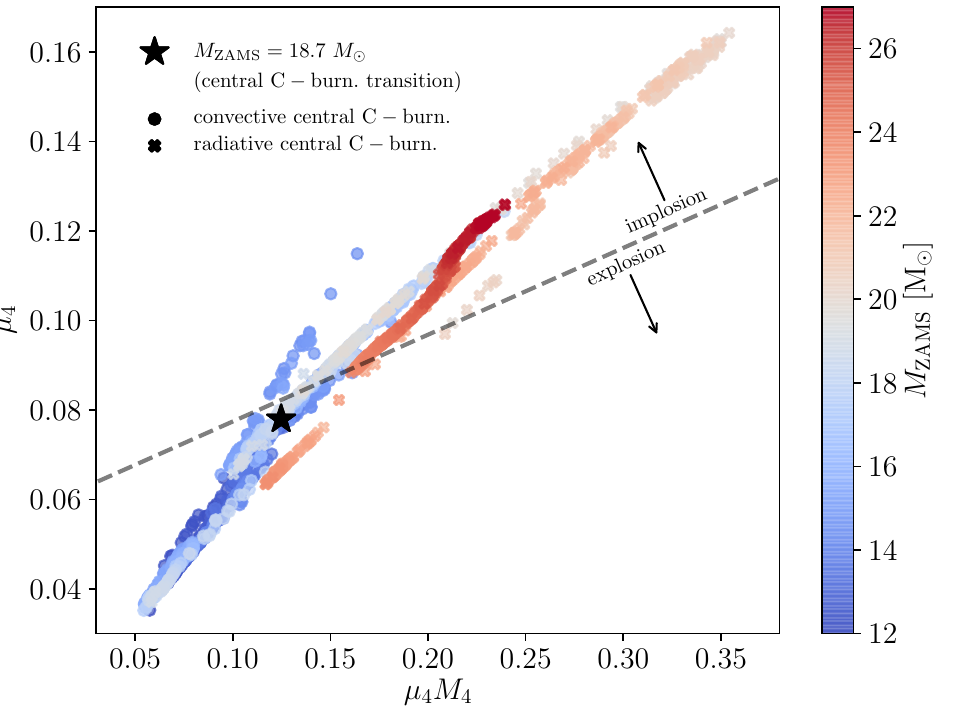}
\caption{Same set of models as in \Fig{f2} are shown on the plane of Ertl-parameters. The dashed line separating explosions from implosions is based on the N20 engine calibration. Stellar models are color coded by their birth mass, and the model corresponding to central Carbon burning transition is denoted by a black star-symbol ($M_{\rm ZAMS}=18.7\ \Msun$), located very close to the critical line. Stars with convective Carbon burning core (circles) are mostly located below the line (explosion), while stars with radiative central Carbon burning (crosses) are found mostly above it (implosion).  \lFig{f3}}
\end{figure}

Not all models above the transition robustly implode, and not all below it explode. As the migration of the innermost Carbon burning shell brings its base outside the effective Chandrasekhar mass in the core, it modulates the size of the Oxygen burning core causing it to ignite earlier with a smaller extent \citep{Suk14}. This results in a narrow island of explosions near $M_{\rm ZAMS}\sim23$ \Msun. Some models right after the transition also manage to explode, while the innermost Carbon burning shell is just starting to migrate out and while $M_4$ is still low. Furthermore, many models before the transition exhibit modulating outcomes due to the possible multi-valued nature of the advanced stages of evolution \citep{Suk18}. Nevertheless, the initial mass space where most stars are likely to explode (below) and implode (above) are clearly delineated by the mass where the central Carbon burning transitions from convective to radiative regime (\Fig{f3}). By number, assuming a Salpeter-IMF with $\alpha=-2.35$, 89\% of RSG between 9 \Msun\ and 18.7 \Msun\ explode, and 84\% between 18.7 \Msun\ and 27 \Msun\ implode. Therefore the observed threshold of exploding RSG progenitors can be viewed as an indirect signature of this critical milestone in massive stellar evolution.

While we demonstrate these arguments using \KEPLER\ calculations, we note that the general result on the connection between this transition mass and the final core structure is evident in models by other codes and groups. Both full-star and bare CO-core calculations by the open source code \MESA\ \citep{Pax11} were in good agreement with that of \KEPLER\ \citep[][see their Figs. 23 and 26]{Suk14}. Despite a very sparse set, it is also apparent from the recent models by the \FRANEC\ code \citep[][see their Fig.22]{Lim18}. The compactness parameter sharply increases (becomes difficult to blow up) after the transition mass, which was about $\sim26$ \Msun\ and $\sim14$ \Msun\ for the non-rotating and $\rm v=150\ km\ s^{-1}$ models respectively (see also discussion in \Sect{constr}).

%%%%%%%%%%%%%%%%%%%%%%%%%%%%%%%%%%%%%%%%%%%%%%%%%%%%%%%%%%%%%%%%%%%%%%%%%%%
\section{Constraining Stellar Models}
\lSect{constr}

The arguments presented in \Sect{trans} suggest that the central Carbon burning transition marks the initial mass point where lighter model stars that experience convective central Carbon burning are intrinsically easier to explode, and heavier stars with radiative central Carbon burning are more likely to implode. Current direct imaging studies indicate a threshold initial mass of about $20$ \Msun, above which RSG do not appear to die in a supernova explosion. Therefore, if we believe the initial mass estimates of progenitors and the current calibrated neutrino-driven explosion models to be reasonably reliable, this presents us an opportunity to constrain massive stellar evolution calculations.

The predictive power of massive stellar evolution models suffer from a number of uncertain input physics, including convective and semi-convective mixing, mass loss, rotation and nuclear reaction rates \citep[e.g.,][]{Ren17,Fie18}. The operation of these uncertainties before the central Carbon ignition can significantly affect the starting mass and composition of the CO-core, and thus the corresponding initial mass for the central Carbon burning transition. While we do not attempt to provide a full systematic study of the dependence of this transition mass on all of the relevant uncertain input physics (currently being explored by Petermann et al., in prep.), we present a general outline of its sensitivity, and two sample sets of calculations demonstrating its dependence on the reaction rate of $^{12}$C$(\alpha,\gamma)^{16}$O, and on the overshoot mixing efficiency. We also survey the transition mass in recently published models by various groups and codes, and discuss how these could potentially be constrained by the initial mass estimates from direct imaging studies.

Convection physics, and in particular mixing at the convective boundaries, remain one of the critical missing components of stellar models \citep[e.g.,][]{Kup17}. While various approaches are being actively investigated \citep[e.g.,][]{Arn15,Gab18}, essentially all existing 1D stellar evolution codes treat the extent of the convective boundary as a free parameter. The relevant structural effect is that Hydrogen burns in a more massive convective core during the main sequence evolution with stronger overshoot mixing, and this ultimately results in more massive embedded He- and CO-cores. In more massive cores the available fuel at central Carbon ignition is also reduced, and therefore the initial mass corresponding to the Carbon burning transition decreases. This general trend is illustrated in \Tab{ov}, which lists the results from a set of \KEPLER\ models with varying overshooting efficiency. For each overshooting efficiency, 60 models were computed between 15 and 21 \Msun\ using the input physics of the standard mass loss ($\dot{M}$) set from \citet{Suk18}. Given the unique implementation of overshooting in the \KEPLER\ code \citep[see \S4.1 of ][]{Suk14}, its efficiency is expressed through an average He-core mass, where a larger the average He-core mass means stronger overshooting. All quantities in \Tab{ov}, excluding the transition mass, are averaged over 10 models between 15$\leq M_{\rm ZAMS}<$16 \Msun, since these can vary substantially between models of nearly identical initial mass. 

\begin{table}
\caption{Effect of Overshooting}
\begin{threeparttable}[b]
\begin{tabular}{lccc}
\hline
$\bar{M}_\alpha$ & $\bar{M}_{\rm CO}$ & $\bar{X}_C(0)$ & Transition Mass\\
$[{\rm M}_{\sun}]$ & $[{\rm M}_{\sun}]$ & & $[{\rm M}_{\sun}]$ \\
\hline
4.37 & 2.79 & 0.247 & 20.2\\
4.48 & 2.81 & 0.223 & 19.1\\
4.56 & 2.86 & 0.214 & 18.7\\
4.63 & 2.91 & 0.208 & 18.2\\
\hline
\lTab{ov}
\end{tabular}
\begin{tablenotes}
\item Note -- overshooting is expressed through the average He-core mass, $\bar{M}_\alpha$. Averaged quantities are evaluated over 10 models between 15$\leq M_{\rm ZAMS}<$16 \Msun\ when the central temperature exceeds ${\rm log}\ T=8.7$ K. The third entry, with the transition mass of 18.7 \Msun, corresponds to the adopted overshooting configuration in the models presented in \Sect{intro} and \Sect{trans}.
\end{tablenotes}
\end{threeparttable}
\end{table}

The effect of rotation also manifests in a similar way, at least for moderately rotating stars. As noted by many prior studies \citep[e.g.,][]{Heg05,Woo02}, rotationally induced mixing results in a more massive He-core for a given initial mass, and a more massive CO-core for a given He-core mass. Calculations from \citet{Chi13} indicate that a $M_{\rm ZAMS}=15$ \Msun\ model rotating with an initial equatorial velocity of 300 $\rm km\ s^{-1}$ has $M_\alpha=5.37$, $M_{\rm CO}=3.59$ and $X_{\rm C}(0)\approx0.18$, while the same model without rotation has $M_\alpha=4.97$, $M_{\rm CO}=2.56$ and $X_{\rm C}(0)\approx0.36$. Therefore, as in the case of overshooting, stronger rotation will lead to a lower initial mass for central Carbon burning transition. However, it should be noted that this general argument does not encompass effects of rapid rotation, such as deformation and enhanced mass loss.

The importance of the  $^{12}$C$(\alpha,\gamma)^{16}$O reaction to nucleosynthesis and massive stellar evolution has long been known \citep[e.g.,][]{Imb01,Heg02,Tur07,Tur09,Wes13}.
Earlier indirect deductions from nucleosynthesis \citep{Wea93} and from ionized interstellar gas \citep{Gar97} were remarkably close to relatively modern rates suggested by \citet[][$S$(300 keV) = 146 keV barn]{Buc06}. However, many challenges such as inconsistencies in current measurements are impeding further reduction of the current $\approx20\%$ uncertainty in the extrapolation for $S$(300 keV) \citep[e.g.,][]{Deb17}. \Fig{f3b} illustrates the sensitivity of the initial mass for central Carbon burning transition on the rate of $^{12}$C$(\alpha,\gamma)^{16}$O. The reference rate used here is that of \citet[][B96]{Buc96}. With higher rates, there is less available fuel for central Carbon burning, and therefore the transition happens at lower initial mass. For the employed input physics in this set of models, the approximate $1\sigma$ variations in the rates from \citet{Deb17} and \citet{Sch12} imply that the transition mass could be as low as 18 \Msun\ or as high as 22 \Msun.

Beside the reactions taking place before the ignition of Carbon, the transition mass will also be highly sensitive to the fusion cross-section of heavy-ion reaction $^{12}$C+$^{12}$C. Recent cross section measurements \citep{Tum18} hint at the possibility of low-energy resonances that could increase the rate by more than factor of 20 (at $5\times10^8$ K) with respect to the standard rate \citep{Cau88}. As pointed out by \citet{Ben12}, higher rates will cause ignition at lower temperatures and thus the transition will happen at a lower initial mass.

\begin{figure}%{r}{0.5\textwidth}
\includegraphics[width=.48\textwidth]{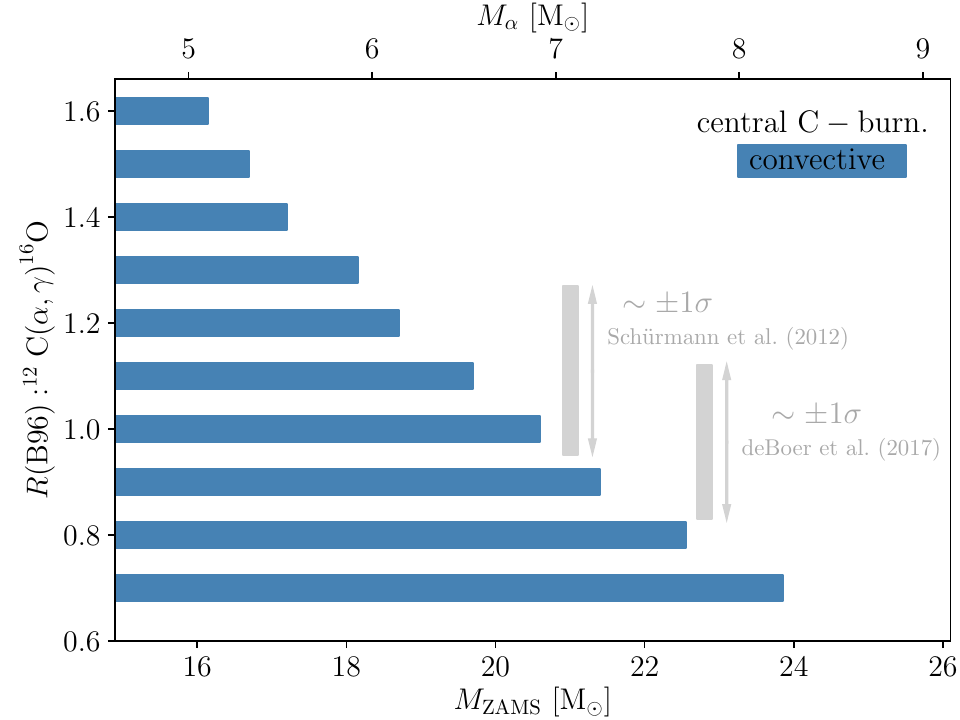}
\caption{Sensitivity of the Carbon burning transition initial mass to the rate of the reaction $^{12}$C$(\alpha,\gamma)^{16}$O. The reference rate, B96, is from \citet{Buc96}, and blue regions denote the models in which Carbon burns in a convective episode at the center. Higher rates reduce the available fuel at Carbon ignition and thus the transition to radiative regime happens at lower initial mass. Shaded gray vertical bars mark the approximate 1$\sigma$ variations based on recently suggested rates. Corresponding He-core masses, $M_\alpha$, are shown on top. \lFig{f3b}}
\end{figure}

The sensitivity to mass loss depends on the adopted prescription. Roughly speaking, the initial mass for the Carbon burning transition will be largely insensitive to algorithms that do not affect the resulting He- and CO-core masses for $M_{\rm ZAMS}\leq 25-30$ \Msun.  For instance, all three sets of models in \citet{Suk18} with varying efficiencies for the \citet{NdJ90} algorithm have the same transition mass near 18.5 \Msun. However, if the algorithm is taking into account the dust formation, rotational enhancement, or bi-stability jump in opacity, the embedded core masses are appreciably affected \citep[e.g., ][]{Ren17,Lim18}, and thus the transition mass for Carbon burning will also be affected.

In \Fig{f4} we have compiled the initial masses corresponding to the Carbon burning transition in various sets of recently published massive star models computed by three different codes. Only solar metallicity models are considered, and the inclusion of rotation is noted. Except in the few cases where we are able to precisely determine the transition mass, we provide ranges, since the model grids are typically very sparse. These ranges are bound by the most massive model with a convective central Carbon burning episode, and the least massive model with radiative burning. 

In the older models computed with the \KEPLER\ code \citep{Suk14,Woo07,Woo02} the transition mass was just above 20 \Msun, while in the newer models \citep{Suk18,Mul16} it has decreased to about 19 \Msun. The slight downward shift is a consequence of updated neutrino losses, which were somewhat underestimated in earlier calculations.

The outcomes from the open source code \MESA\ \citep{Pax11} depend on the version of the code and input configurations. Using the version \emph{r4930} and similar input physics, \citet{Suk14} were able to largely recreate the \KEPLER\ results, with a transition mass of roughly 19 \Msun. However, the models recently computed by \citet{Rit18} using a similar version \emph{r3709} but with different input physics, find a much larger transition initial mass. In their solar metallicity 20 \Msun\ model, Carbon burns convectively in the core with an extent of $\sim$ 0.4 \Msun, which implies the transition mass happening near $\sim$25 \Msun. The input physics were substantially different, including in convective overshooting efficiency and the rate of $^{12}$C$(\alpha,\gamma)^{16}$O.

Carbon burns convectively in a small core (0.04 \Msun) in the non-rotating 25 \Msun\ \FRANEC\ model from \citet{Chi13}, and thus the transition mass in this set was probably close to $\sim26$ \Msun. For the rotating set (with 150 $\rm km\ s^{-1}$) the transition mass was much smaller, between 15 and 20 \Msun. In the more recent set by \citet{Lim18}, the transition mass was also just above 25 \Msun\ for the non-rotating case, however, with rotation (also 150 $\rm km\ s^{-1}$) it is decreased to about $\sim$14 \Msun. It should be noted here that the adopted mass loss was so strong that only non-rotating models with $<20$ \Msun\ died as RSG, while heavier models, and all rotating models (even at $M_{\rm ZAMS}=13\ \Msun$) died as compact blue stars.  

These comparisons illustrate that different calculations from various groups produce vastly different transition masses, ranging from 14 to 26 \Msun. While the range inferred from currently available direct imaging constraints is not precise, $16\lesssim M_{\rm ZAMS}\leq20$ \Msun\ \citep{Sma15}, it is much narrower than the spread seen in these different sets of calculations. Given the importance of massive stellar models in many astrophysical problems, we encourage stellar modelers to consider tuning the uncertain input physics such that the Carbon burning transition is achieved roughly between 15 and 20 \Msun\ at solar metallicity. %$16<M_{\rm ZAMS}<20$ \Msun.

%%%%%%%%%%%%%%%%%%%%%%%%%%%%%%%%%%%%%%%%%%%%%%%%%%%%%%%%%%%%%%%%%%%%%%%%%%%
\section{Caveats}
\lSect{disc}

The arguments presented in this paper are certainly not ironclad, and one needs to consider number of caveats from both observational and theoretical sides.

For instance, \citet{Wal12} argued that the initial mass estimates from direct imaging studies do not take into account circumstellar extinction resulting from the dust produced in RSG winds \citep[see also ][]{Bea16}. According to their analysis, the true progenitor luminosities and masses are grossly underestimated in direct imaging studies, such that the ``missing RSG problem'' would effectively not exist. However, \citet{Koc12} pointed out that this effect has to be small since the circumstellar dust can scatter photons from the central star into the line-of-sight as well as out of it -- scattering has little net effect on the observed flux. Moreover, the dust composition from a single star can differ dramatically from the mixture observed in the interstellar medium.  Thus, they also argued that modelling circumstellar dust with an interstellar extinction law is incorrect, and it inherently overestimates the amount of extinction, and hence the luminosities and masses of progenitors.

\begin{figure}%{r}{0.5\textwidth}
\includegraphics[width=.48\textwidth]{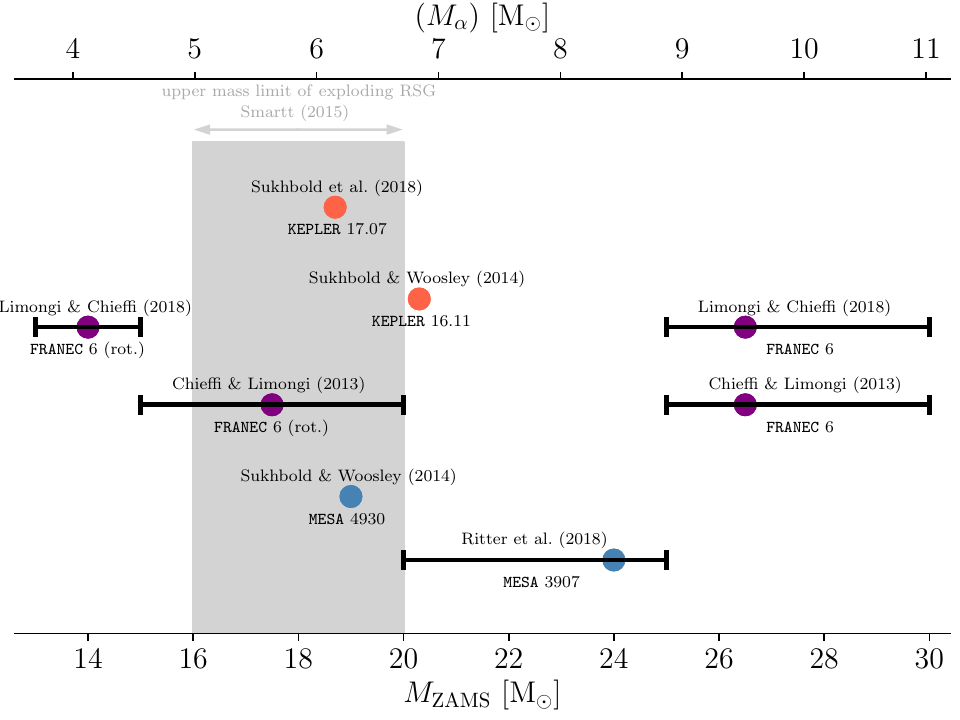}
\caption{The diversity of initial masses corresponding to the Central Carbon burning transition in various recently published solar metallicity massive star models. The sets that included rotation are indicated with ``rot.''. In most cases the mass increments between models were larger than a solar mass, and these cases are represented as a range bounded by the largest mass with central convective episode and the smallest mass with radiative burning. The central values (circles) in the sparse sets are estimated based on the mass extent in the most massive model with convective central burning (see text). The gray vertical band represents the upper mass limit of exploding RSG inferred from direct imaging studies \citep{Sma15}. The He-core masses corresponding to the models of \citet{Suk18} is shown on top, however, note that the relation $M_{\rm ZAMS}\propto M_\alpha$ will vary depending on the input physics and numerical implementations across different codes. We suggest stellar modelers to tune the uncertainties such that the transition is achieved near or within this observed range, which is roughly $16<M_{\rm ZAMS}<20$ \Msun, or $5<M_\alpha<6.8$ \Msun, or $5.07<\log L/\Lsun<5.26$.  \lFig{f4}}
\end{figure}

More recently, \citet{Dav18} questioned the evidence for the missing luminous RSG by arguing that the bolometric corrections used to convert pre-explosion flux to bolometric luminosity are not properly treated in earlier studies. Using empirically determined bolometric corrections they re-analyzed a subset of 24 events, and report an upper mass limit of $\sim25$ \Msun, which implies that there is no statistically significant difference between the observed highest mass of a RSG and the upper limit inferred from progenitor imaging. However, their analysis is biased towards higher mass as they employed 3$\sigma$ (in one case 5$\sigma$) values for upper limits rather than 1$\sigma$.  In a future work (Kochanek, in prep) the claims of \citet{Dav18} will be examined through detailed Monte-Carlo simulations.

From the theoretical side, a lot of the model outcomes discussed in this work heavily rely on the complicated interplay of advanced stage episodes and the resulting non-monotonic final core structures as a function of initial mass as seen in \KEPLER\ calculations. While this general result was recreated with another code \citep[using \MESA, ][]{Suk14}, it is not clear if other groups are seeing this. Part of the problem is that some codes and groups do not evolve the stars past Carbon burning, and others publish only a very limited number of models in this important mass range. For example, the recent study with \FRANEC\ by \citet{Lim18} had only 5 models across the critical initial mass range between $8<M_{\rm ZAMS}<30$ \Msun, as compared to more than 1000 in \citet{Suk18}. It would be a fruitful future exercise to perform a detailed comparative study on the advanced stage evolution and final core structures involving all major stellar evolution codes.

While the final fate of the star firmly depends on the presupernova core structure, the outcome is ultimately determined by the model for the explosions. The results discussed in this study were based on parameterized neutrino-driven explosion calculations by \citet{Ert16}, which employed a version of the code \PHOTB. These results are in good overall agreement with parameterized calculations by other codes such as \GRD\ \citep[e.g.,][]{Pej15,Ott11}, and in excellent agreement with an independently developed semi-analytical model by \citet{Mul16}. However, it is in tension with a recently proposed approach based on 1D turbulence model by \citet[][see also \citealp{Mab19}]{Cou19}, in which they find many stars above 20 \Msun\ exploding as supernova and many imploding below it. This result is not only inconsistent with various observational constraints, but it has also been argued that energy conservation is violated in these types of approaches \citep{Mul19}. Finally, we also note that the explosion modeling discussed in this work is limited to neutrino-driven scenario \citep[e.g.,][]{Jan16}. It has been suggested that higher mass stars may retain significant amount of angular momentum in their cores \citep[e.g.,][]{Heg05}, and likely the neutrino-driven scenario is not the full story. There is no census yet in the community on how the explosion landscape is altered by magneto-rotational effects.

With all these caveats in mind, it is also worth highlighting the potentially testable prediction made by the stellar evolution and explosion models discussed in this study --  the ``islands'' of implosions below the upper mass limit of exploding RSG, and ``islands'' of explosions above the threshold (\Fig{map}). Although, a detection of stellar implosion, perhaps near 15 \Msun, would be strong evidence to support this scenario, it is a challenging task unlikely to happen in the near future. It is observationally expensive to search for the disappearance of a RSG in nearby galaxies.  The first seven years of the LBT survey \citep{Koc08} monitoring a million RSGs in 27 nearby galaxies only yielded one such event \citep{Ada17b}.  This approach cannot easily be scaled up.  Even LSST will only increase the failed SN discovery rate by $\sim$50\% since the LBT survey already monitors 2/3 of the $M_{R}<-6$ RSGs within 11 Mpc.  Only a large, multi-year time-domain survey of galaxies within 30$-$50 Mpc with WFIRST would dramatically increase the rate.
% mention Dave Sand's HST program? https://ui.adsabs.harvard.edu/#abs/2018hst..prop15645S/abstract 
% The search for disappearing stars in archival \emph{HST} data \citep{Rey15} could be expanded with new \emph{HST} observations...

An alternative approach is to search for the weak transients predicted to be associated with failed SNe.  The faint but long-lived transient from the ejected hydrogen envelope  appears to have been observed for the failed SN candidate of \citet{Ger15} and \citet{Ada17}.  At $10^{6}~L_{\odot}$, such a transient is too faint to be found by current SN surveys and even when detecting such a transient becomes possible with LSST it will be difficult to impossible to distinguish whether the transient is a failed SN or an unrelated weak transient (such as stellar merger).  However, in a failed SN the long-lived faint transient should be preceded by a few-day $10^{7}~L_{\odot}$ shock breakout that radiates a large fraction of its energy in the optical \citep{Pir13,Lov17,Fer18}.  With new SN surveys it should be possible to detect the failed SN shock breakout, which can then trigger follow-up spectroscopy and later deep imaging (to detect the fainter, long-lived transient component).  With the high-cadence survey of ZTF the expected yield is roughly $0.5$--$1~\mathrm{yr}^{-1}$ \citealt{Gra19}), but not all implosions may result in these weak transients and while the observables of the luminosities, temperatures, and timescales of the fast shock breakout phase and the longer-lived, fainter recombination-powered transient can constrain the explosion energy, ejected mass, and progenitor radius, they do not provide strong constraints on the progenitor mass.

However, testing the existence of an explosion ``island'' above the current upper mass limit is perhaps possible in the next decade. As the Carbon burning shells migrate out with increasing initial mass after the transition, it temporarily modulates the central Oxygen burning episode as it passes through the effective Chandrasekhar mass in the core, and creates a relatively narrow mass range near $\sim23$ \Msun\ where stars become easier to blow up again. Adopting the model outcomes as presented in the bottom section of \Fig{map}, and assuming a Salpeter IMF with $\alpha=-2.35$ covering the range between 8 and 30 \Msun, the fraction occupied by the explosion ``island'' near $M_{\rm ZAMS}\sim23$ \Msun\ is about 0.02--0.03 of all successful supernovae. Currently we have 29 combined detections and upper limits, that increases by about 1.5 per year for Type II events  \citep[e.g.,][]{Van17}. Therefore, in the next decade we may be able to find convincing evidence to support or rule out this feature. While the lack of this explosion ``island'' would not spell a complete doom to this scenario, it will certainly force us to rework some of the key pieces of stellar and core-collapse supernova physics.

%%%%%%%%%%%%%%%%%%%%%%%%%%%%%%%%%%%%%%%%%%%%%%%%%%%%%%%%%%%%%%%%%%%%%%%%%%%
\section{Conclusions}

In this study we explore the connection between the critical transition of central Carbon burning from convective to radiative regime in massive stars, and the initial mass upper limit inferred from direct imaging studies of Type IIP supernova progenitors. Our key conclusions are:
\begin{itemize}

\item Currently available progenitor imaging studies strongly support the existence of upper initial mass limit of 16 -- 20 \Msun, where heavier RSG stars do not experience supernova explosion at the end of their life. This result, also known as the ``missing RSG problem'', is in a very good agreement with recent calibrated neutrino-driven explosion results based on progenitor models that carefully followed the advanced stage evolution of massive stars.

\item These models suggest that the observed upper mass limit of exploding RSG stars is innately tied to the critical transition of central Carbon burning episode in massive stars from convective to radiative regime. In lighter stars Carbon burns convectively in the center ($M_{\rm ZAMS}\lesssim 20$ \Msun), which causes Carbon burning shells to operate deep in the core and prevents the development of massive central Oxygen burning episode. The resulting presupernova core structures are typically very compact, meaning lower mass iron-core surrounded by steeply declining density profile, which tend to be easy to blow up. In heavier stars, after the transition, Carbon burns radiatively in the center, and causes the Carbon burning shells to rapidly migrate outward with increasing initial mass. This allows the development of massive Oxygen burning cores and extended final core structures that are generally much more difficult to blow up. The central Carbon burning transition mass neatly delineates the initial mass range where most stars are expected to explode, from the range where most stars are expected to implode.

\item The central Carbon burning transition mass is sensitive to number of key uncertain input physics of massive stellar evolution. We demonstrate its sensitivity to the rate of the reaction $^{12}$C$(\alpha,\gamma)^{16}$O, the efficiency of overshoot mixing, and argue that it must also be sensitive to stellar rotation, mass loss, and other key reaction rates and convective physics uncertainties.

\item Recently published massive stellar evolution models from various groups and codes exhibit wide ranging initial masses corresponding to the Carbon burning transition. Some are as low as $\sim14$ \Msun, while others are as high as $\sim26$ \Msun. We suggest stellar modelers consider tuning their model uncertainties so that the central Carbon burning transition takes place roughly within or near the range inferred from direct imaging studies, 16 -- 20 \Msun, or roughly  $5.07<\log L/\Lsun<5.26$.

\item There are potential caveats from both observational and theoretical sides. The direct imaging measurements could be biased, the RSG problem is still based on a small number of measurements, and the models are yet to be extensively tested by other codes and methods. However, the models discussed in this study predict an ``island'' of explosion near $\sim23$ \Msun, just above the observed mass limit. The existence of this feature could be tested through a larger sample of direct imaging measurements in the next decade.

\end{itemize}

%%%%%%%%%%%%%%%%%%%%%%%%%%%%%%%%%%%%%%%%%%%%%%%%%%%%%%%%%%%%%%%%%%%%%%%%%%%
\section*{Acknowledgements}

We thank the referee for many valuable comments that helped to improve the paper. We also would like to thank Stan Woosley, Christopher Kochanek, Thomas Janka, Alexander Heger, and Marco Limongi for valuable comments, and Falk Herwig for sharing the details of their calculations. All numerical \KEPLER\ calculations presented in \Sect{constr} were performed on the \texttt{RUBY} cluster at the Ohio Supercomputer Center \citep{osc}. Support for this work was provided by NASA through the NASA Hubble Fellowship grant \#60065868 awarded by the Space Telescope Science Institute, which is operated by the Association of Universities for Research in Astronomy, Inc., for NASA, under contract NAS5-26555.

\Leg{Software:} \texttt{matplotlib} \citep{Hun07}, \texttt{numpy} \citep{Van11}.
%%%%%%%%%%%%%%%%%%%%%%%%%%%%%%%%%%%%%%%%%%%%%%%%%%

%%%%%%%%%%%%%%%%%%%% REFERENCES %%%%%%%%%%%%%%%%%%

%%%%%%%%%%%%%%%%%%%%%%%%%%%%%%%%%%%%%%%%%%%%%%%%%%

% Don't change these lines
\bsp	% typesetting comment
\label{lastpage}
\end{document}